\documentstyle[12pt,a4]{article}

\begin{document}
\title{Analysis by neutron activation analysis a some ancient Dacian ceramics}
\author{Agata Olariu\\
{\em National Institute for Physics and Nuclear Engineering}\\
{\em P.O.Box MG-6, 76900 Bucharest Magurele, Romania}\\}

\maketitle

\section{Introduction}

Ceramics is the most common archaeological material and therefore it is 
very used material by the historians to draw temporal and cultural
characterizations. The importance of knowledge the compositional scheme of 
the pottery is well known$^{1-5}$ although very rarely one can get important
conclusions from the elemental analysis of the potsherds$^{6,8}$.
Perlman and Assaro$^{7}$, on the basis of an analysis of thousands of objects 
of ancient ceramics by neutron activation
analysis established a method of classification on the objects in well defined 
groups, characterized on the historic point of view (culture, dating, 
style, etc).   \\
In this paper we have been analyzed by neutron activation analysis (NAA)
samples of ancient Dacian ceramics, from Romanian territories, from 3 
different establishments from Romanian territory: Strei San Giorgiu, Hunedoara, 
Popesti, Giurgiu and Fierbinti, Ialomita. Ceramics was delivered by National 
Museum of History from Bucharest. We have searched the characteristic element 
or the ratio of elements for a given Dacian archaeological settlement.

\section{Experimental method}

The samples listed in the Table 1 have been analyzed by neutron activation
analysis. 
We considered that the analysis should give an image of the bulk of the objects
and therefore the surface of the shards was removed. Also we had into
consideration the homogeneity of the samples and that the samples must 
be representative for the whole object.
Samples of potsherds have been cut, weighted and wrapped 
individually in plastic foil. The samples of 10-30 mg have been irradiated at
at the rabbit system of VVR-S reactor of the NIPNE,
Bucharest-Magurele, at a flux of 1.3 x 10$^{12}$ 
neutrons/cm$^2$$\cdot$sec$^{-1}$, for a period of 15 minutes. A standard 
spectroscopic pure metallic
copper was used as neutron flux standard. 

The measurements have been performed
with a Ge(Li) detector, 135 cm$^3$ coupled at PC with a MCA interface. The
system gave a resolution of 2.4 keV at 1.33 MeV ($^{60}$Co). 
The radioactivity of the samples has
been measured after a decay of $^{27}$Al (T$_{1/2}$=2.54 min), the major 
element in the structure of
ceramics. Then after a decay time of 10 min we have observed in the $\gamma$
spectra of the irradiated samples $\gamma$ rays of the isotopes corresponding
to the elements: Ba, Mn and Na. After a decay
time of 3-4 d we measured again the radioactivity in the ceramics samples and we 
could observed the elements: Sm, Eu, Sc, La and K.

\section{Results and discussions}

In the Table 2, 3 and 4 are shown the results of the NAA
of the three groups of ancient Dacian potsherds S, P and F, from the three 
different establishments: Strei San Giorgiu, Hunedoara, Popesti, 
Giurgiu and Fierbinti, Ialomita.
The concentrations are given in ppm, and when the concentration was larger than
10,000 ppm the result was given in percents. The considered statistical errors 
have been $\approx$1\% for Mn and Na and $<$5\% for the others elements.\\
One could observe from the Fig. 1 that the concentration of Ba seams to vary
relatively  from one group to other.
The mean values and the standard deviations for the concentration of Ba for
each group are the following:
\begin{center}
\begin{tabular}{cc}
 Ceramics  Strei San Giorgiu & C$_{Ba}$=2248$\pm$833 ppm\\
 Ceramics Popesti              & C$_{Ba}$=796$\pm$226 ppm \\
 Ceramics Fierbinti             & C$_{Ba}$=4138$\pm$467 ppm \\
\end{tabular}
\end{center}
Then we applied the procedure to consider the ratio Na/Mn 
found to be constant in ancient ceramics, for a given archaeological settlement,
 for Maya period$^{4}$.\\
The values of the means and the standard deviations of the ratio Na/Mn, 
for the 3 groups of
analyzed Dacian ceramics are the following:
\begin{center}
\begin{tabular}{cc}
 Ceramics  Strei San Giorgiu & Na/Mn=28.3$\pm$22.9 ppm\\
 Ceramics  Popesti              & Na/Mn=7.40$\pm$3.53 ppm \\
 Ceramics Fierbinti             & Na/Mn=7.92$\pm$1.68 ppm \\
\end{tabular}
\end{center}
We could remove in calculus of the means, the values of the concentrations 
far away of the mean  and one get the following values:
\begin{center}
\begin{tabular}{cc}
 Ceramics  Strei San Giorgiu & Na/Mn=15.04$\pm$0.07 ppm\\
 Ceramics  Popesti              & Na/Mn=5.91$\pm$1.30 ppm \\
 Ceramics Fierbinti             & Na/Mn=7.32$\pm$1.18ppm \\
\end{tabular}
\end{center}
For the case of analyzed samples of ancient Dacian potsherds we could say that 
the ratio of the concentrations of Na/Mn is not constant and can not
characterize a given settlement. Ba is the elements that could be considered
to relatively differentiate the three groups of ceramics.  
To draw conclusions it must to improve the statistics
of analysis and also to pay more attention to the homogeneity of the samples,
that in the case of ceramics it is a very important parameter of the analysis.

\newpage

{\bf\large References}

\noindent
1. A. Aspinal, D. N. Slater, "Neutron activation analysis of medieval 
ceramics", Nature 217 (1968) 368\\
2. J. S. Olin and Ed. V. Sayre, "Trace analysis of English and American pottery
of the american colonial period" The 1968 Intern. Conference of Modern Trends
in Activation Analysis" (1968) p. 207\\ 
3. N. Saleh, A. Hallak and C. Bennet, "PIXE analysis of ancient ordanian
pottery", Nuclear Instruments and Methods 181 (1981) p. 527\\
4. Ed. Sayre, "Activation Analysis applications in art and archaeology", in
Advances in Activation Analysis, eds. J.M.A. Lenihan, S.J. Thomson and V.P.
Guinn, Academic Press, London, p.157\\
5. Ch. Lahanier, F.D. Preusser and L. Van Zelst, "Study and conservation of
museum objects: use of classical analytical techniques", Nuclear Instruments
and Methods, B14 (1986) p.2\\
6. Zvi Goffer, Archaeological Chemistry, Chemical Analysis, Vol. 55, eds. P.J.
Elving J. D. Winefordner, John Wiley \& Sons, p.108\\
7. I. Perlman and F. Assaro, "Deduction of provenience of pottery from trace
element analysis", Scientific Methods in Medieval Archaeology, ed. R. Berger
Univ. of California Press (1970) p.389\\
8. A. Millet and H. Catling, "Composition and provenance: a challenge",
Archaeometry, Vol. 9 (1966) p.92\\

\newpage

\begin{table}

\newlabel{}
\caption{{\bf Table 1}. List of analyzed ancient Dacian potsherds }\\

\begin{tabular}{ccc}
\hline

Sample  & Period & Source\\
\hline
\hline\\
S1, S2, S3    & cent. I B.C.- I. A.D. & Strei San Giorgiu, Hunedoara\\
              &                          &                   \\
P1, P2, P3, P4, P5 & cent. I B.C. & Popesti, Giurgiu\\
              &                          &                   \\
F1, F2, F3, F4, F5 & cent. IV - III B.C. & Fierbinti, Ialomita\\ 
              &                          &                   \\
\hline
\end{tabular}
\end{table}

\begin{table}
\newlabel{}
\caption{{\bf Table 2}. Ancient ceramics from Strei San Georgiu, concentrations in
ppm, by NAA}\\

\begin{tabular}{ccccccccc}
\hline
Sample  & Sm & Eu & Ba & La & Mn & Sc & Na & Na/Mn\\
\hline
\hline\\
S1    & 7.54  & - & 2910 & 42.8 & 599 & -  & 8969 & 14.9\\
S2    & 7.13  & - & 2520 & 109  & 336 & 8.1& 18400& (54.8)\\
S3    &  7.74 & - & 1313 & 21.9 & 388 & 9.1 & 5864 & 15.1\\
\hline
\end{tabular}
\end{table}

\begin{table}
\newlabel{}
\caption{{\bf Table 3}. Ancient ceramics from Popesti, concentrations in
ppm, by NAA}\\

\begin{tabular}{ccccccccc}
\hline
Sample  & Sm & Eu & Ba & La & Mn & Sc & Na & Na/Mn\\
\hline
\hline\\
P1    & 8.8  & - & 953 & 48.3 & 703 & 23.3  & 4488 & 6.4\\
P2    & 7.1  & - & 749 & 44.0 & 888 & 19.6  & 6497 & 7.3\\
P3    & 7.5 & - & 505  & 43.1 & 90  & 11.0  & 1205 & 14.1\\
P4    & 8.6 & - & 1083 & 50.8 & 690 & 11.4  & 3940 & 5.7\\
P5    & 9.7 & - & 692  & 47.3 & 789 & 13.6  & 333  & 4.5\\
\hline
\end{tabular}
\end{table}

\newpage
\begin{table}
\newlabel{}
\caption{{\bf Table 4}. Ancient ceramics from Fierbinti, concentrations in
ppm, by NAA}\\

\begin{tabular}[h]{ccccccccc}
\hline
Sample  & Sm   & Eu   & Ba & La & Mn & Sc & Na & Na/Mn\\
\hline
\hline\\
F1    & 10.6  & 0.95 & 4110 & 51.1 & 495 & 16.9  & 3803 & 7.7\\
F2    & 8.8   & 1.2  & 5860 & 40.3 & 503 & 24.2  & 4036 & 8.0\\
F3    & 8.1   & 1.0  & 3150 & 30.7 & 767 & 16.9  & 4269 & 5.6\\
F4    & 7.2   & 0.91 & 4070 & 55.6 & 488 & 18.6  & 5032 & 10.3\\
F5    & 7.6   & 1.02 & 3500 & 30.3 & 663 & 17.9  & 5303 & 8.0\\
\hline
\end{tabular}
\end{table}

\end{document}